\documentclass{article}
\usepackage{amssymb}
\usepackage{graphicx}
\usepackage{amsmath}

\input{tcilatex}

\begin{document}

\title{Scaling of Decoherence Effects in Quantum Computers}
\author{B. J. DALTON \dag \ddag  \\
\dag Centre for Atom Optics and Ultrafast Spectroscopy,\\
Swinburne University, Hawthorn, Victoria 3122, \\
Australia\\
\ddag\ Department of Physics, University of Queensland,\\
St Lucia, Queensland 4072, Australia}
\maketitle

\begin{abstract}
The scaling of decoherence rates with qubit number N is studied for a simple
model of a quantum computer in the situation where N is large. The two state
qubits are localised around well-separated positions via trapping potentials
and vibrational centre of mass motion of the qubits occurs. Coherent one and
two qubit gating processes are controlled by external classical fields and
facilitated by a cavity mode ancilla. Decoherence due to qubit coupling to a
bath of spontaneous modes, cavity decay modes and to the vibrational modes
is treated. A non-Markovian treatment of the short time behaviour of the
fidelity is presented, and expressions for the characteristic decoherence
time scales obtained for the case where the qubit/cavity mode ancilla is in
a pure state and the baths are in thermal states. Specific results are given
for the case where the cavity mode is in the vacuum state and gating
processes are absent and the qubits are in (a) the Hadamard state (b) the
GHZ state.
\end{abstract}

\section{$\subset $Introduction}

Quantum computers are expected to have a key advantage over classical
computers in terms of computational complexity, an advantage based on the
idealised features of parallelism, entanglement and unitary quantum
computation processes. For certain algorithms \cite{Shor95a, Grover97}, the
number of computational steps needed using quantum computers should increase
much more slowly with input number size than for classical computers,
enabling computations that were infeasible using classical computers to be
carried out. In the case of the quantum search algorithm for example, the
number of steps increases with the square root of the number of items,
rather than linearly as in classical searching \cite{Grover97}. However, the
physical system whose quantum states define the N qubit system and the
quantum devices involved in the gating processes both interact with the
environment. Such interactions change the density operator describing the
quantum computer state from a pure to a mixed state, with coherences between
states evolved from different input states being partially or completely
destroyed. This process of decoherence is the enemy of quantum computation,
though methods such as quantum error correction \cite{Shor95b, Steane96},
decoherence free sub-spaces \cite{Lidar98, Plenio99, Beige00a, Jane02} and
dynamical suppression of decoherence \cite{Viola98, Gea-Banacloche01} could
be used to minimize its effects. Utilizing the computational complexity
advantage of quantum computers requires suitably large qubit numbers (ca $%
10^{5}$ qubits may be needed for searching or factoring algorithms with
error correction \cite{Preskill99} for input numbers of practical
importance), so the scaling with the number of qubits of the decoherence
rates in comparison to the coherent processing rates is important in
determining the size limits of useful quantum computers \cite{Plenio96,
Plenio97}.

This paper extends previous work \cite{Unruh95, Palma96, Garg96, Duan98a,
Duan98b, Duan98c, Sorensen00} treating scaling effects. Decoherence effects
are studied for a simple model of an $N$ qubit quantum computer, involving $N
$ two state qubit systems (see figure 1). The qubits are localised around
well-separated positions via trapping potentials, and the centre of mass
(CM) vibrational motions of the qubits are treated. Coherent one and two
qubit gating processes are controlled by time dependent localised classical
electromagnetic (EM) fields and magnetic fields that address specific
qubits. The two qubit gating processes are facilitated by a cavity mode
ancilla, which permits state interchange between qubits. The magnetic fields
are used to bring specific qubits into resonance with the classical EM
fields or the cavity mode. The two state qubits are coupled to a bath of EM
field spontaneous emission (SE) modes, and the cavity mode is coupled to a
bath of cavity decay modes. For large $N$ the numerous vibrational modes of
the qubits also act as a reservoir, coupled to the qubits, the cavity mode
and the SE modes. The system-environment coupling interactions include
electric dipole coupling of the qubits to the SE modes, quasi-mode coupling
of the cavity mode to cavity decay EM\ field modes and Lamb-Dicke coupling
of qubits to CM vibrational modes and each of: (a) the SE modes (b) the
cavity mode (c) the classical EM gating fields (d) the classical magnetic
gating field. The system-environment coupling interactions considered
include both $\sigma _{Z}$ phase destroying and $\sigma _{X}$ population
destroying terms, though the latter are more important in this model. The
model is similar to those of the ion-trap \cite{Cirac95}, neutral atom \cite%
{Briegel00} and cavity QED \cite{Domokos95, Pellizzari95, Turchette95}
varieties.

The object is to study decoherence effects in quantum computers for the
situation where the number of qubits $N$ becomes large. Decoherence effects
will be specified by the fidelity, which measures how close is the actual
behaviour of the density operator for the qubits system (including the
ancilla) to its idealised behaviour due to coherent gating processes only.
The initial quantum state for the quantum computer will be assumed to be
pure and the reservoir states will be assumed to be thermal. Previous work 
\cite{Unruh95, Palma96, Garg96, Duan98a, Duan98b, Duan98c, Sorensen00}
indicates that the decoherence time $t_{D}$ can decrease inversely with $N$
(qubits decohering independently) or $N^{2}$ (qubits decohering
collectively), so it can become very short for the case where $N$ is large
and may become comparable to the reservoir correlation time $t_{C}$. In the
case of atomic spontaneous emission relaxation, $t_{C}$ $\sim 10^{-17}$ s,
and $t_{D}$ $\sim 10^{-8}$s for one qubit could become $t_{D}$ $\sim
10^{-18} $s for $10^{5}$ qubits in the case of collective decoherence.
However, one aspect of the present study is whether the qubits decohere
independently or not, so we do not arbitrarily assume either case. As
maintaining coherence between all states of the quantum computer is
important for its operation, we are interested in the behaviour of \textit{%
all} density matrix elements, and not just those for a smaller density
matrix describing a single qubit, even though the latter may have long
decoherence times and satisfy Markovian equations. Treating the density
operator for the full $N$ qubit system via the standard Born-Markoff master
equation (see for example, \cite{Barnett97, Dalton99}) requires that \textit{%
all} the interaction picture density matrix elements do not change
significantly during the reservoir correlation times. As the latter are just
the time scales over which two-time correlation functions of reservoir
operators involved in the system-reservoir interaction decay to zero, we can
see that the reservoir correlation times are independent of the number of
qubits. Since the general decrease in decoherence time with increasing qubit
numbers indicates that \textit{some }density matrix elements must change
over shorter and shorter time scales, then using Markovian master equations
for studying decoherence therefore becomes questionable if the number of
qubits becomes large enough In the present paper we do not assume that
Markovian behaviour will necessarily occur, and use methods that do not rely
on the standard Born-Markoff master equation. If the behaviour does turn out
to be Markovian then our approach is still correct. However as we will see,
in certain cases the behaviour is not consistent with the Markoff
approximation, so the necessity for our choice of non-Markovian methods is
justified a posteriori. Several authors \cite{Unruh89, Braun01, Privman02,
Strunz02} have also emphasised the importance of studying the short time
regime for macroscopic and mesoscopic systems (such as quantum computers
with large qubit numbers) using non-Markovian methods. One such approach 
\cite{Braun01, Privman02} involves using eigenstates of the system operators
involved in the system-reservoir interaction (pointer basis), another
treatment \cite{Retamal01} develops short-time expressions for the linear
entropy (or impurity). Here the approach used is similar to that in \cite%
{Duan97} and uses the Liouville-von Neumann equations to obtain short-time
expansions for the fidelity. This leads to expressions for the
characteristic decoherence time scales for the general case with gating EM\
and magnetic fields present, and at non-zero temperature, and with the qubit
and ancilla system in an arbitrary pure state. The treatment enables
situations where independent or collective decoherence occur to be
distinguished. In this initial paper, the decoherence time scales will be
evaluated for specific initial quantum computer states such as: (a) the
equal superposition of all qubit states (Hadamard state) and (b) the
generalised GHZ states, with the cavity mode ancilla in the vacuum state and
gating processes ignored. The effects of gating processes and the
relationship of the results to those obtainable for longer times using
Markovian master equation methods will be treated later.

In Section 2 the general theory used will be outlined. Results for the
decoherence time scales are presented in Section 3 and Section 4 contains a
summary of the paper.

\section{Theory}

\subsection{Hamiltonian}

The total Hamiltonian is written as the sum: 
\begin{equation}
H=H_{S}+H_{C}+H_{B}+V_{S}+V_{I}
\end{equation}

The Hamiltonian for the qubit system and cavity mode ancilla is given in
terms of qubit atomic spin operators $\sigma _{Z}^{i}$ and cavity mode
annihilation, creation operators $b,b^{\dagger }$ by: 
\begin{equation}
H_{S}=\sum_{i}\frac{1}{2}\hbar \omega _{0}\sigma _{Z}^{i}+\hbar \omega
_{b}b^{\dag }b
\end{equation}%
Here the qubits all have transition frequency $\omega _{0}$ and $\omega _{b}$
is the cavity mode frequency. The cavity mode facilitates state transfer
between different qubits in two qubit gating process. In the case of a large
number of charged qubits, the CM vibrational modes are too closely spaced to
enable the in-phase vibrational mode to be used in this role, as in the case
of ion-trap quantum computers \cite{Cirac95}. The array of qubits confined
in their trapping potentials is rather analogous to a low density solid.
Also, because the cavity mode energy quantum is large, this ancilla is not
subject to thermal effects as are the low frequency CM vibrational modes.

The Hamiltonian for the centre of mass vibrational motion of the qubit
system is given in the harmonic approximation as: 
\begin{eqnarray}
H_{C} &=&\frac{1}{2m}\sum_{i\alpha }p_{i\alpha }^{2}+\frac{1}{2}%
\sum_{ij\alpha \beta }V_{ij}^{\alpha \beta }\delta r_{i\alpha }\delta
r_{j\beta } \\
&=&\sum_{K}\hbar \nu _{K}A_{K}^{\dag }A_{K}
\end{eqnarray}%
In these equations the qubit CM momentum operators are $p_{i\alpha }$ and
the centre of mass of each qubit vibrates with displacements $\delta
r_{i\alpha }$ about well-separated positions $\mathbf{r}_{i0}$\ defined by
trapping potentials $(\alpha ,\beta =x,y,z)$. The vibrational mode $K$ has
angular frequency $\nu _{K}$ and phonon annihilation, creation operators $%
A_{K},A_{K}^{\dag }$. The vibrational Hamiltonian $H_{C}$ applies to the
general case where the qubits vibrate collectively, such as in the ion-trap
case where the charged qubits interact via long-range Coulomb potentials. If
the qubits vibrate independently, such as for neutral qubits with
dipole-dipole couplings ignored, then the coupling matrix is given by $%
V_{ij}^{\alpha \beta }=\delta _{ij}\delta _{\alpha \beta }V$, and the
vibrational modes $K$ are specified $i\alpha $. These all have the same
frequency $\nu $.

The Hamiltonian for the bath of spontaneous emission and cavity decay modes
is of the form: 
\begin{equation}
H_{B}=\sum_{k}\hbar \omega _{k}a_{k}^{\dag }a_{k}+\sum_{k}\hbar \xi
_{k}b_{k}^{\dag }b_{k}
\end{equation}%
Here the SE mode $k$ has angular frequency $\omega _{k}$ and annihilation,
creation operators $a_{k},a_{k}^{\dagger }$, and the cavity decay mode $k$
has angular frequency $\xi _{k}$ and annihilation, creation operators $%
b_{k},b_{k}^{\dagger }$. These modes are described in the quasi-mode picture 
\cite{Dalton99b}.

The coherent coupling (or gating) processes for the qubit system are
described by the Hamiltonian: 
\begin{eqnarray}
V_{S} &=&\sum_{i}(\hbar \Omega _{i}\sigma _{X}^{i}+Hc)+\sum_{i}\frac{1}{2}%
\hbar (\Delta _{i1}-\Delta _{i0})\sigma _{Z}^{i}  \notag \\
&&+\sum_{i}(\hbar g_{i}\sigma _{X}^{i}b+Hc)
\end{eqnarray}%
In order, the terms are the coupling of the qubits to the classical EM\
field, to the magnetic field and to the cavity mode. Each qubit can be
addressed by localised classical EM fields $\mathbf{E}$ and magnetic fields $%
\mathbf{B}$ to facilitate one and two qubit gating processes. The two qubit
gating process involves the creation and annihilation of a single photon in
the cavity mode, which acts as an ancilla. The quantities $\Omega
_{i},\Delta _{ia}$and $g_{i}$ are coupling constants. The coupling constants 
$\Omega _{i},\Delta _{ia}$ are both spatially localised and time dependent
in accordance with the sequence of gating processes involved for a
particular algorithm.

The interaction of the qubit system and ancilla with the baths and the
qubits centre of mass vibrational degrees of freedom is given by: 
\begin{eqnarray}
V_{I} &=&\sum_{ik}(\hbar g_{k}^{i}\sigma _{X}^{i}a_{k}+Hc)+\sum_{k}(\hbar
b^{\dag }(u_{k}b_{k}+w_{k}b_{k}^{\dag })+Hc)  \notag \\
&&+\sum_{ikK}(\hbar n_{kK}^{i}\sigma _{X}^{i}a_{k}[A_{K}+A_{K}^{\dag }]+Hc) 
\notag \\
&&+\sum_{iK}(\hbar p_{K}^{i}\sigma _{X}^{i}b[A_{K}+A_{K}^{\dag }]+Hc)  \notag
\\
&&+\sum_{iK}(\hbar \Theta _{K}^{i}\sigma _{X}^{i}[A_{K}+A_{K}^{\dag }]+Hc) 
\notag \\
&&+\sum_{iK}\frac{1}{2}\hbar (m_{0K}^{i}(1-\sigma
_{Z}^{i})+m_{1K}^{i}(1+\sigma _{Z}^{i}))[A_{K}+A_{K}^{\dag }]
\end{eqnarray}%
The first two terms are the electric dipole coupling of the qubits to SE
modes and the quasi-mode coupling of the cavity mode to the cavity decay
modes. These terms involve coupling constants $g_{k}^{i},u_{k}$ and $w_{k}$.
The remaining terms are Lamb-Dicke couplings with the various fields,
allowing for the vibrational motion of the qubits CM around their
equilibrium positions $\mathbf{r}_{i0}$. The third term allows for coupling
to the SE modes, the fourth to the cavity mode, the fifth to the classical
EM field and the sixth to the magnetic field. These terms involve coupling
constants $n_{kK}^{i},p_{K}^{i},\Theta _{K}^{i}$ and $m_{0K}^{i}$. Since
this paper studies short time scale effects, the standard rotating wave
approximation (RWA) has \textit{not} been made. Other possible interactions,
such as those due to Roentgen currents \cite{Baxter93, Wilkens94} for
neutral and charged qubit systems, and those due to ionic currents \cite%
{Baxter93, Wilkens94} for charged qubit systems, were also investigated but
found to be relatively small in comparison to the interactions listed above.

\subsubsection{Expressions}

Detailed definitions or commutation rules for the qubit, cavity mode, CM
vibrational modes and bath modes operators are as follows: 
\begin{eqnarray}
\sigma _{X}^{i} &=&(|1\rangle \langle 0|+|0\rangle \langle 1|)_{i}\quad
\sigma _{Y}^{i}=-i(|1\rangle \langle 0|-|0\rangle \langle 1|)_{i}  \notag \\
\sigma _{Z}^{i} &=&(|1\rangle \langle 1|-|0\rangle \langle 0|)_{i} \\
\lbrack b,b^{\dag }] &=&1\quad \lbrack A_{K},A_{L}^{\dag }]=\delta _{KL} 
\notag \\
\lbrack a_{k},a_{l}^{\dag }] &=&\delta _{kl}\quad \lbrack b_{k},b_{k}^{\dag
}]=\delta _{kl}
\end{eqnarray}%
The qubits are represented by Pauli spin operators, whereas bosonic
annihilation and creation operators apply for the cavity mode (quantized in
volume $V_{b}$), the CM vibrational modes, the SE modes and the cavity decay
modes (quantized in volume $V$).

The CM displacements $\delta r_{i\alpha }$ $(\alpha =x,y,z)$ are related to
the vibrational normal coordinates via a unitary matrix $S$ in the form: 
\begin{equation}
\delta r_{i\alpha }=\sum_{K}S_{i\alpha ;K}\sqrt{\frac{\hbar }{2m\nu _{K}}}%
(A_{K}+A_{K}^{\dag }),
\end{equation}%
where the matrix $S$ is determined from the eigenvalue equations 
\begin{equation}
\sum_{j\beta }V_{ij}^{\alpha \beta }S_{j\beta ;K}=m\nu _{K}^{2}S_{i\alpha
;K}.
\end{equation}%
These equations also give the vibrational frequencies $\nu _{K}$. For the
case where the qubits vibrate independently $S_{i\alpha ;K}=\delta _{i\alpha
;K}$ and $\nu =(V/m)^{\frac{1}{2}}$. 

Explicit expressions for the coupling constants are: 
\begin{eqnarray}
\Omega _{i} &=&-i\sum_{c}\sqrt{\frac{\omega _{c}}{2\epsilon _{0}\hbar V}}(%
\mathbf{d}_{10}\cdot \mathbf{e}_{c})\alpha _{c}\exp i(\mathbf{k}_{c}\cdot 
\mathbf{r}_{i0}-\omega _{c}t) \\
\Delta _{ia} &=&-[\mathbf{m}_{aa}\cdot \mathbf{B]}/\mathbf{\hbar \quad \quad 
}a=0,1 \\
g_{i} &=&-i\sqrt{\frac{\omega _{b}}{2\epsilon _{0}\hbar V_{b}}}(\mathbf{d}%
_{10}\cdot \mathbf{e}_{b})\exp (i\mathbf{k}_{b}\cdot \mathbf{r}_{i0}) \\
g_{k}^{i} &=&-i\sqrt{\frac{\omega _{k}}{2\epsilon _{0}\hbar V}}(\mathbf{d}%
_{10}\cdot \mathbf{e}_{k})\exp (i\mathbf{k}\cdot \mathbf{r}_{i0}) \\
n_{kK}^{i} &=&\sqrt{\frac{\omega _{k}}{2\epsilon _{0}\hbar V}}\sqrt{\frac{%
\hbar }{2m\nu _{K}}}(\mathbf{d}_{10}\cdot \mathbf{e}_{k})(\mathbf{k}\cdot 
\mathbf{S}_{iK})\times  \notag \\
&&\times \exp (i\mathbf{k}\cdot \mathbf{r}_{i0}) \\
p_{K}^{i} &=&\sqrt{\frac{\omega _{b}}{2\epsilon _{0}\hbar V_{b}}}\sqrt{\frac{%
\hbar }{2m\nu _{K}}}(\mathbf{d}_{10}\cdot \mathbf{e}_{b})(\mathbf{k}%
_{b}\cdot \mathbf{S}_{iK})\times  \notag \\
&&\times \exp (i\mathbf{k}_{b}\cdot \mathbf{r}_{i0}) \\
\Theta _{K}^{i} &=&\sum_{c}\sqrt{\frac{\omega _{c}}{2\epsilon _{0}\hbar V}}%
\sqrt{\frac{\hbar }{2m\nu _{K}}}(\mathbf{d}_{10}\cdot \mathbf{e}_{c})(%
\mathbf{k}_{c}\cdot \mathbf{S}_{iK})\times  \notag \\
&&\times \alpha _{c}\exp i(\mathbf{k}_{c}\cdot \mathbf{r}_{i0}-\omega _{c}t)
\\
m_{aK}^{i} &=&-\sum_{\beta }{\Large (}\frac{\partial (\mathbf{m}_{aa}\cdot 
\mathbf{B)}}{\partial R_{i\beta }}{\Large )}_{0}\sqrt{\frac{\hbar }{2m\nu
_{K}}}S_{i\beta ;K}\;/\mathbf{\hbar \quad }a=0,1.
\end{eqnarray}%
Here $\mathbf{d}_{10}$ is the electric dipole matrix element, $\mathbf{m}%
_{aa}(a=0,1)$ are the magnetic dipole matrix elements for each qubit. The
polarisation unit vectors for the classical field, the cavity mode, the SE
mode or cavity decay mode $k$ are $\mathbf{e}_{c},\mathbf{e}_{b},\mathbf{e}%
_{k}$ respectively, and $\mathbf{k}_{c},\mathbf{k}_{b},\mathbf{k}$ give the
wave vectors. The classical EM gating field could be thought of as being
represented by a multi-mode coherent state, where $\alpha _{c}$ is the
coherent state amplitude for the mode $c$. All the modes have essentially
the same polarisation vector, wave vector and frequency, though the
frequency and wave vector bandwidths are sufficiently wide to produce the
required spatio-temporal localisation. The equivalent classical electric
field at the equilibrium position of the $i$ qubit is given by $\mathbf{E}%
=i\sum_{c}\sqrt{\frac{\hbar \omega _{c}}{2\epsilon _{0}V}}\mathbf{e}%
_{c}\alpha _{c}\exp i(\mathbf{k}_{c}\cdot \mathbf{r}_{i0}-\omega _{c}t)+cc.$

\bigskip

\subsection{Dynamics}

The total density operator $W$ for the full system of qubits, qubits CM
motion, cavity mode and baths satisfies the Liouville-von Neumann equation: 
\begin{equation}
i\hbar \frac{\partial W}{\partial t}=[H,W]
\end{equation}%
Initial conditions are assumed of an uncorrelated state for qubits and
ancilla, bath and CM vibrational motion given by:. 
\begin{equation}
W(0)=\rho _{S}(0)\rho _{B}(0)\rho _{C}(0)
\end{equation}%
and for the bath and CM in thermal states the average value of $V_{I}$ is
zero. 
\begin{equation}
Tr_{BC}V_{I}\rho _{B}(0)\rho _{C}(0)=Tr_{BC}\rho _{B}(0)\rho _{C}(0)V_{I}=0
\end{equation}%
In the present paper the initial state for the qubits and ancilla will be
assumed in a pure state $|\psi _{S}\rangle $, so that $\rho _{S}(0)=|\psi
_{S}\rangle \langle \psi _{S}|$.

The dynamics of the quantum computer is described by the reduced density
operator $\rho _{S}$ for the qubits and ancilla subsystem defined by: 
\begin{equation}
\rho _{S}=Tr_{BC}W
\end{equation}%
The exact evolution allowing for the coherent coupling and the interaction
with the bath and the CM vibrational modes is embodied in the time
dependence of $\rho _{S}$. No assumption is made here that $\rho _{S}$
satisfies a Markovian master equation, though it may do so for time scales
which are large compared to bath correlation times.

The coherent evolution of the qubits and ancilla subsystem is described by
the reduced density operator $\rho _{S0}$, which is chosen to coincide with
the exact reduced density operator at time zero, and is then allowed to
evolve under the coherent coupling term $V_{S}$ only. Hence: 
\begin{eqnarray}
i\hbar \frac{\partial \rho _{S0}}{\partial t} &=&[H_{S}+V_{S},\rho _{S0}] \\
\rho _{S0}(0) &=&\rho _{S}(0)
\end{eqnarray}%
Thus the interaction between the qubits/ancilla and the baths and CM motion
is considered to be switched off here, hence $\rho _{S0}$ evolves according
to a Liouville-von Neumann equation..

\subsection{Decoherence\ }

Decoherence effects are specified by the fidelity, defined as: 
\begin{equation}
F=Tr_{S}(\rho _{S0}\rho _{S})
\end{equation}%
The fidelity specifies how close the actual evolution of qubits and ancilla
is to the idealised coherent evolution. The time dependence of the fidelity
is entirely due to the decoherence effects caused by the bath and CM
interactions with the qubits and cavity mode ancilla.

As in \cite{Duan97} the short-time behaviour of the fidelity with the qubits
in a pure state can be expressed as a power series in the time elapsed and
explicit expressions obtained for the first few time constants involved.
Thus: 
\begin{eqnarray}
F(t) &=&1-(\frac{t}{\tau _{1}})-(\frac{t^{2}}{2\tau _{2}^{2}})+.. \\
\frac{\hbar }{\tau _{1}} &=&0 \\
\frac{\hbar ^{2}}{2\tau _{2}^{2}} &=&Tr_{BC}\langle \psi
_{S}|V_{I}(0)^{2}|\psi _{S}\rangle \rho _{B}(0)\rho _{C}(0)  \notag \\
&&-Tr_{BC}\langle \psi _{S}|V_{I}(0)|\psi _{S}\rangle ^{2}\rho _{B}(0)\rho
_{C}(0) \\
&\equiv &\langle \langle \Delta V_{I}(0)^{2}\rangle _{S}\rangle _{BC}
\end{eqnarray}%
The initial value of the fidelity is unity. The times $\tau _{1},\tau
_{2},.. $specify characteristic decoherence times for the qubit and ancilla
system, their inverses defining decoherence rates. For thermal bath and CM
states, only $\tau _{2}$ is involved in specifying short time decoherence.
For the time constant $\tau _{2}$, the expression involves the average of
the square of the fluctuation $\Delta V_{I}(0)=V_{I}(0)-\langle
V_{I}(0)\rangle _{S}$ of the zero time interaction operator $V_{I}(0)$. The
squared fluctuation operator is first averaged over the initial qubit and
ancilla pure state $|\psi _{S}\rangle $, then the result is averaged over
the bath and CM initial state $\rho _{B}(0)\rho _{C}(0)$.

\section{Results}

\subsection{General case}

The evaluation of $\tau _{2}$\ for the general case with gating EM\ and
magnetic fields both present, and at non-zero temperature, and with the
qubit and ancilla system in an arbitrary pure state, leads to expressions
for $\tau _{2}$\ that involve:

(a) Expectation values for the state $|\psi _{S}\rangle $ of one qubit
operators $\sigma _{\alpha }^{i}$, two qubit operators $\sigma _{\alpha
}^{i}\sigma _{\beta }^{j}\;(i\neq j),$ $(\alpha ,\beta =X,Z)$.

(b) Expectation values of cavity operators $b,b^{\dag },bb^{\dag },b^{\dag
}b,b^{2},b^{\dag 2}$.

(c) Expectation values of products of these qubit and cavity operators.

(d) Sums over bath modes and centre of mass vibrational modes of products of
pairs of the coupling constants $g_{k}^{i},w_{k},n_{kK}^{i},p_{K}^{i},\Theta
_{K}^{i},m_{aK}^{i},$ the products being weighted by factors involving the
thermally averaged photon and phonon quantum numbers $\overline{n}_{k},%
\overline{m}_{k},\overline{N}_{K}$ for the SE modes, cavity decay modes and
vibrational modes. The thermally averaged quantum numbers are given by the
Planck function.

The general case will be studied in further work.

\subsection{Case\ of spontaneous emission for stationary qubits, no cavity
mode}

In this case the cavity mode is ignored, the qubits are assumed stationary
so that CM\ vibrational modes are excluded (as are all Lamb-Dicke
interactions), and the only coupling constants included are $g_{k}^{i}$,
given by $g_{k}\exp ik\cdot r_{i0}$. The qubit pure state is $|\phi
_{Q}\rangle $. This special situation has been studied in \cite{Duan97},
where temperature effects and criteria for the qubits decohering
independently or collectively were examined.

For this case the decoherence time $\tau _{2}$ is given by:%
\begin{eqnarray}
\frac{1}{2\tau _{2}^{2}} &=&\sum_{ij}(\langle \sigma _{X}^{i}\sigma
_{X}^{j}\rangle -\langle \sigma _{X}^{i}\rangle \langle \sigma
_{X}^{j}\rangle )\sum_{k}g_{k}^{i}\,g_{k}^{j\ast }(2\overline{n}_{k}+1) \\
&=&\sum_{ij}\langle \Delta \sigma _{X}^{i}\Delta \sigma _{X}^{j}\rangle
\sum_{k}|g_{k}|^{2}\cos \mathbf{k\cdot d}_{ij}\coth \frac{\hbar \omega _{k}}{%
2k_{B}T}
\end{eqnarray}%
where $\langle ..\rangle =\langle \phi _{Q}|..|\phi _{Q}\rangle ,\Delta
\Omega =\Omega -\langle \Omega \rangle $\ and $\mathbf{d}_{ij}=\mathbf{r}%
_{i0}-\mathbf{r}_{j0}$. This result is the same as in \cite{Duan97}.

As in \cite{Duan97}, if $\Delta k$\ is the bandwidth and $\overline{k}$\ the
mean of a normalised Gaussian model for the function $|g_{k}|^{2}\coth \frac{%
\hbar \omega _{k}}{2k_{B}T}$,\ then the condition for independent
decoherence (where only terms with $i=j$\ contribute to $\tau _{2}$)
requires that $\Delta k\,d_{ij}\gg 1$ for any pair of qubits $%
(d_{ij}=\left\vert \mathbf{d}_{ij}\right\vert )$. Collective decoherence
(where terms from all pairs $i\neq j$\ of qubits contribute to $\tau _{2}$)
requires that $\Delta k\,d_{ij}\ll 1$ and $\overline{k}$\thinspace $%
d_{ij}\ll 1$. Further discussion about independent and collective
decoherence is given below for the case of zero temperature, but based on
evaluating the sum over SE modes $k$ for a three dimensional model.

\subsection{Case of no gating processes, zero temperature and cavity in
vacuum state}

The case where no gating is taking place, the ancilla is back in its
original no photon state $|0\rangle _{A}$ and the temperature is zero,
provides a simple case of the general results, enabling the dependence on
the qubit state to be examined. Here $|\psi _{S}\rangle =|\phi _{Q}\rangle
|0\rangle _{A}$, where the state of the qubit system is $|\phi _{Q}\rangle $%
. This case is of some physical interest, since it corresponds to states
produced after idealised coherent gating processes have occurred. Too rapid
a decoherence for such states would be of concern for the implementation of
quantum computers.

\subsubsection{Case of correlated qubit states}

\begin{eqnarray}
\frac{1}{2\tau _{2}^{2}} &=&\sum_{ij}(\langle \sigma _{X}^{i}\sigma
_{X}^{j}\rangle -\langle \sigma _{X}^{i}\rangle \langle \sigma
_{X}^{j}\rangle )G_{ij}  \notag \\
&&+\sum_{ij}\langle \sigma _{X}^{i}\sigma _{X}^{j}\rangle K_{ij}+L \\
G_{ij} &=&\sum_{k}g_{k}^{i}\,g_{k}^{j\ast
}+\sum_{kK}n_{kK}^{i}\,n_{kK}^{j\ast } \\
K_{ij} &=&\sum_{K}p_{K}^{i}\,p_{K}^{j\ast }\quad \\
L &=&\sum_{k}w_{k}\,w_{k}^{\ast }
\end{eqnarray}%
For the correlated qubits case, the decoherence time depends on SE coupling
constants, non-RWA cavity mode coupling constants and Lamb-Dicke coupling
constants involving the cavity mode and the SE modes. Note that terms
involving the same qubits $(i=j)$ and different $(i\neq j)$ qubits are
involved.

\subsubsection{Case of uncorrelated qubit states}

In this case $\langle \sigma _{X}^{i}\sigma _{X}^{j}\rangle =\langle \sigma
_{X}^{i}\rangle \langle \sigma _{X}^{j}\rangle $ for $i\neq j$,\ giving the
decoherence time via:{\small \ } 
\begin{eqnarray}
\frac{1}{2\tau _{2}^{2}} &=&\sum_{i}(1-|\langle \sigma _{X}^{i}\rangle
|^{2})(G_{ii}+K_{ii})  \notag \\
&&+\sum_{ij}\langle \sigma _{X}^{i}\rangle \langle \sigma _{X}^{j}\rangle
^{\ast }K_{ij}+L \\
&=&\sum_{i}(1-|\langle \sigma _{X}^{i}\rangle |^{2})(\sum_{k}\left\vert
g_{k}^{i}\right\vert ^{2}+\sum_{kK}\left\vert n_{kK}^{i}\right\vert
^{2}+\sum_{K}\left\vert p_{K}^{i}\right\vert ^{2})  \notag \\
&&+\left\vert \sum_{i}\langle \sigma _{X}^{i}\rangle
\sum_{K}p_{K}^{i}\right\vert ^{2}+L
\end{eqnarray}%
For the uncorrelated qubits case, the decoherence time depends on the same
coupling constants as for the correlated case, but now only diagonal terms
involving the same qubits are involved.

\subsubsection{Special sub-case of Hadamard or fiducial state}

This particular uncorrelated state results from applying the Hadamard one
qubit gating process to every qubit, and which produces the important
fiducial state of an equal superposition of states representing all numbers
from $0$ to $2^{N}-1$. The state vector for the $i$ qubit is $|\phi
_{Q}\rangle _{i}=(|0\rangle _{i}+|1\rangle _{i})/2^{1/2}$. In this case the
decoherence time is given by: 
\begin{eqnarray}
\frac{1}{2\tau _{2}^{2}} &=&\sum_{ijK}p_{K}^{i}p_{K}^{j\ast
}+\sum_{k}|w_{k}|^{2} \\
\sum_{ijK}p_{K}^{i}p_{K}^{j\ast } &\thickapprox &N\eta ^{2}g_{b}^{2}
\label{LambDicke1.eq}
\end{eqnarray}%
{\small \ }The first term is associated with the characteristic time scale
due to the Lamb-Dicke term coupling the qubits, cavity mode and vibrational
motion. For the general case of collective qubit vibrations, this term is
only evaluated approximately, the $1/\nu _{K}$ factor appearing in the sum
over $K$ being replaced by an average value, and thus enabling the unitary
properties of the matrix $S$ to be employed. Only $i=j$ terms then result.
For the case of independent qubit vibrations, result (\ref{LambDicke1.eq})
is exact. The Lamb-Dicke parameter $\eta \thickapprox \mathbf{k}\cdot 
\mathbf{\delta r}$\ is small compared to unity. $g_{b}\thickapprox (\frac{%
\omega _{b}}{2\epsilon _{0}\hbar V_{b}})^{1/2}\left\vert \mathbf{d}%
_{10}\right\vert $ is the cavity vacuum Rabi frequency. The second term is
associated with the characteristic time scale due to the non-RWA term
coupling the cavity mode and cavity decay modes. Ignoring this term, the
decoherence time scales as $1/\sqrt{N}$ for the Hadamard state. For $%
g_{b}\thickapprox 50$ Mhz, $\eta \thickapprox 10^{-2}$\ and thus $\eta
g_{b}\thickapprox 10^{6}$s$^{-1}$,\ we find that the decoherence time scale
is about $10^{-8}$s for $N\thickapprox 10^{4}$\ qubits. If the quantum
computer were to become truly macroscopic with $N\thickapprox 10^{22}$, the
decoherence time is about $10^{-17}$s. Decoherence effects associated with
the SE modes do not occur for this state, as can also been seen from the
expressions for the case of stationary qubits.

\subsubsection{Special sub-case of GHZ\ state}

This correlated state is of special interest due to its highly entangled
nature. The state vector for the qubit system is $|\phi _{Q}\rangle
=(|00..0\rangle +|11..1\rangle )/2^{1/2}$and the decoherence time obtained
from:

\begin{eqnarray}
\frac{1}{2\tau _{2}^{2}} &=&\sum_{ik}|g_{k}^{i}|^{2}+%
\sum_{ikK}|n_{kK}^{i}|^{2}  \notag \\
&&+\sum_{iK}|p_{K}^{i}|^{2}+\sum_{k}|w_{k}|^{2} \\
\sum_{ik}|g_{k}^{i}|^{2} &=&N\sum_{k}|g_{k}|^{2} \\
\sum_{ikK}|n_{kK}^{i}|^{2} &\thickapprox &N\eta ^{2}\sum_{k}|g_{k}|^{2} \\
\sum_{iK}|p_{K}^{i}|^{2} &\thickapprox &N\eta ^{2}g_{b}^{2}
\end{eqnarray}%
The first term in the expression for $\tau _{2\text{ }}$is associated with
the characteristic time scale due to the electric dipole term coupling the
qubits and SE modes. It can be expressed in terms of the RMS vacuum electric
field $\mathbf{E}_{rms}$\ for the SE modes as proportional to the square of
the Rabi frequency $(\mathbf{E}_{rms}\cdot \mathbf{d}_{10}/\hbar )$\
associated with $\mathbf{E}_{rms}$. The vacuum Rabi frequency for the $k$ SE
mode is $g_{k}\thickapprox (\frac{\omega _{k}}{2\epsilon _{0}\hbar V_{k}}%
)^{1/2}\left\vert \mathbf{d}_{10}\right\vert $. The sum $\sum_{k}|g_{k}|^{2}%
\ $can be evaluated for SE modes in three dimensions by incorporating a
cut-off factor for $|g_{k}|^{2}$ of the form $\exp (-\omega _{k}/\omega _{c})
$, where $\omega _{c}$\ is cut-off frequency. The correlation time for the
bath of SE modes is given by the inverse of the cut-off frequency and is
about 10$^{-17}$s. We find that:{\small \ } 
\begin{equation}
\sum_{k}{\small |g}_{k}{\small |}^{{\small 2}}\thickapprox {\small \Gamma
\omega }_{c}{\small (\omega }_{c}/{\small \omega }_{0}{\small )}^{{\small 3}}
\end{equation}%
Here $\Gamma $ is the SE decay rate. For $\Gamma \thickapprox 10^{8}$s$^{-1}$%
, $\omega _{0}\thickapprox 10^{15}$s$^{-1}$, $\omega _{c}\thickapprox 10^{17}
$s$^{-1}\ $we find that $(\sum_{k}|g_{k}|^{2})^{1/2}\thickapprox 10^{15}$s$%
^{-1}$.The second term in the expression for $\tau _{2\text{ }}$is
associated with the characteristic time scale due to the Lamb-Dicke term
coupling the qubits, the SE modes and the CM vibrational motion. For the
general case of collective qubit vibrations, this term is only evaluated
approximately. For the case of independent qubit vibrations, the result is
exact. As the Lamb-Dicke parameter is small compared to unity, the second
term can be ignored in comparison with the first. The third term in the
expression for $\tau _{2\text{ }}$is associated with the characteristic time
scale due to the Lamb-Dicke term coupling the qubits, the cavity mode and
the CM vibrational motion. For the general case of collective qubit
vibrations, this term is also only evaluated approximately. For the case of
independent qubit vibrations, the result is exact. It involves the square of
the Rabi frequency associated with the cavity mode. For $g_{b}\thickapprox 50
$ Mhz, $\eta \thickapprox 10^{-2}$\ and thus $\eta g_{b}\thickapprox 10^{6}$s%
$^{-1}$, so the third term is also small compared to the first. The fourth
term in the expression for $\tau _{2\text{ }}$is associated with the non-RWA
term coupling the cavity mode and cavity decay modes. Ignoring the last
term, the decoherence time scales as $1/\sqrt{N}$ for the GHZ state. As the
first term is the most important, we see that the decoherence time scale is
about $10^{-17}$s for $N\thickapprox 10^{4}$\ qubits. If the quantum
computer became macroscopic with $N\thickapprox 10^{22}$\ the decoherence
time is even shorter, about $10^{-26}$s. Evidently the GHZ state would not
preserve its entanglement for very long.

\subsubsection{Case of correlated qubit states - spontaneous emission
decoherence only}

The question of whether the qubits decohere collectively or independently is
now examined further. For the case of stationary-qubits at zero temperature
and with no cavity mode ancilla and only the coupling to the SE modes
included, the decoherence time $\tau _{2}$\ is given as a special case of
the previous expression as:

\begin{equation}
\frac{1}{2\tau _{2}^{2}}=\sum_{ij}\langle \Delta \sigma _{X}^{i}\Delta
\sigma _{X}^{j}\rangle \sum_{k}|g_{k}|^{2}\cos \mathbf{k\cdot d}_{ij}
\end{equation}%
Evaluating the sum $\sum_{k}|g_{k}|^{2}\cos \mathbf{k}\cdot \mathbf{d}_{ij}$%
\ for SE modes in three dimensions and incorporating a cut-off factor for $%
|g_{k}|^{2}$ of the form $\exp (-\omega _{k}/\omega _{c})$\ as before, the
sum is given by the expression:{\small \ } 
\begin{equation}
\sum_{k}|g_{k}|^{{\small 2}}\cos \mathbf{k}\cdot \mathbf{d}_{ij}\thickapprox
\Gamma \omega _{c}(\omega _{c}/\omega _{0})^{{\small 3}}F(\omega _{c}\tau
_{ij},\cos ^{{\small 2}}\theta _{ij;10})
\end{equation}%
where $\tau _{ij}=d_{ij}/c$\ is the time for light to travel between the $i$%
\ and $j$\ qubits, $\theta _{ij;10}$\ is the angle between the vectors for
the separation of the qubits $\mathbf{d}_{ij}$\ and the dipole moment $%
\mathbf{d}_{10}$. The function $F$\ is of order unity and involves ratios of
polynomials in the quantity $\omega _{c}\tau _{ij}$. For $\omega _{c}\tau
_{ij}\gg 1$ this function behaves like $(\omega _{c}\tau _{ij})^{-4}$. This
is in contrast to the one dimensional case at low temperatures treated in 
\cite{Duan97}, where no such decrease for large $\omega _{c}\tau _{ij}$\
occurs. For $\Gamma \thickapprox 10^{8}$s$^{-1}$, $\omega _{0}\thickapprox
10^{15}$s$^{-1}$, $\omega _{c}\thickapprox 10^{17}$s$^{-1}\ $(as above) we
find that $(\Gamma \omega _{c}(\omega _{c}/\omega
_{0})^{3})^{1/2}\thickapprox 10^{15}$s$^{-1}$. Allowing for conveniently
addressing separate qubits with optical laser fields, the nearest neighbour
distance will be taken as $d_{ij}\thickapprox 10^{-6}$m, hence $\omega
_{c}\tau _{ij}\thickapprox 10^{3}\gg 1$. Thus even for nearest neighbour
qubits, $(\Gamma \omega _{c}(\omega _{c}/\omega
_{0})^{3}F)^{1/2}\thickapprox 10^{9}$s$^{-1}\ $. This is a factor of $10^{6}$%
\ smaller than for the $i=j$ terms. For more widely separated qubits the
reduction is even greater. The greater number of qubits in a shell of qubits
between $d_{ij},d_{ij}+\delta d_{ij}$\ (proportional to $d_{ij}^{2}$) does
not overcome the $d_{ij}^{-4}$\ factor. Thus it would seem that only the $%
i=j $ terms will be important in the expression for the decoherence time $%
\tau _{2}$. This indicates that the qubits decohere independently. Including
then only the $i=j$ terms (which are all equal), the expression for $1/2\tau
_{2}^{2}$ is proportional to $N$ and is non zero for correlated states.
Overall, the decoherence time scales as $1/\sqrt{N}$ for correlated states.
Similar considerations to before again lead to a decoherence time scale of
about $10^{-17}$s for $N\thickapprox 10^{4}$\ qubits, whilst for a
macroscopic size quantum computer with $N\thickapprox 10^{22}$\ the
decoherence time is about $10^{-26}$s. Remembering that the SE bath
correlation time is of order $\omega _{c}^{-1}\approx 10^{-17}$s, then these
decoherence times are not long compared to the correlation time, so
non-Markovian conditions apply as has been expected. If spontaneous emission
is present this would seem to be something of a problem for quantum
computation based on the model treated here, since processing involving
correlated states is an essential feature.

\subsection{Case of no gating processes, zero temperature, cavity in vacuum
state but spontaneous emission absent}

The case where no gating is taking place, the ancilla is back in its
original no photon state $|0\rangle _{A}$, the temperature is zero, and SE
modes are ignored provides a simple case of the general results, enabling
the dependence on the qubit state to be examined. Here it is assumed that
the cavity is of such high quality that qubit decay via the SE modes can be
ignored, so the only decay is via Lamb-Dicke coupling through the cavity
mode. Again $|\psi _{S}\rangle =|\phi _{Q}\rangle |0\rangle _{A}$, where the
state of the qubit system is $|\phi _{Q}\rangle $. This case is of some
physical interest, since it corresponds to states produced after idealised
coherent gating processes have occurred, but now in a more favourable cavity
situation. Too rapid decoherence of correlated qubit states - even in this
case where SE decay is absent - would be of concern for the implementation
of quantum computers.

\subsubsection{Case of Correlated qubit States}

The previous expressions can be used to obtain the decoherence time $\tau
_{2}$ by deleting all contributions associated with the SE modes. This gives:

\begin{eqnarray}
\frac{1}{2\tau _{2}^{2}} &=&\sum_{ij}\langle \sigma _{X}^{i}\sigma
_{X}^{j}\rangle K_{ij}+L \\
K_{ij} &=&\sum_{K}p_{K}^{i}\,p_{K}^{j\ast }\quad
L=\sum_{k}w_{k}\,w_{k}^{\ast }
\end{eqnarray}

\subsubsection{Evaluation of Lamb-Dicke Term}

For the general case of collective qubit vibrations, an approximate
evaluation of the Lamb-Dicke term yields the result: 
\begin{equation}
\frac{1}{2\tau _{2}^{2}}\thickapprox N\eta ^{2}g_{b}^{2}
\label{LambDicke2.eq}
\end{equation}%
{\small \ }Result (\ref{LambDicke2.eq}) is exact for the case of independent
qubit vibrations. This term is associated with the characteristic time scale
due to the Lamb-Dicke term coupling the qubits, cavity mode and vibrational
motion. The Lamb-Dicke parameter $\eta \thickapprox \mathbf{k}\cdot \mathbf{%
\delta r}$\ is small compared to unity. $g_{b}\thickapprox (\frac{\omega _{b}%
}{2\epsilon _{0}\hbar V_{b}})^{1/2}\left\vert \mathbf{d}_{10}\right\vert $
is the cavity vacuum Rabi frequency. The decoherence time scales as $1/\sqrt{%
N}$ . For $g_{b}\thickapprox 50$ Mhz, $\eta \thickapprox 10^{-2}$\ and thus $%
\eta g_{b}\thickapprox 10^{6}$s$^{-1}$,\ the decoherence time scale is found
to be about $10^{-8}$s for $N\thickapprox 10^{4}$\ qubits, whilst for a
macroscopic sized quantum computer with $N\thickapprox 10^{22}$\ the
decoherence time is about $10^{-17}$s. Although it might be thought that
these decoherence times are rather long compared to bath correlation times,
it should be noted that for the CM vibrational modes bath, the correlation
time is of order $\nu ^{-1}$, where $\nu $ is a typical CM vibrational
frequency. As $\nu \approx 2\pi .10^{6}$ s$^{-1}$ \cite{Cirac95, Briegel00},
the correlation time for the bath is about $10^{-6}$s. Thus these
decoherence times are short compared to the correlation time, so
non-Markovian conditions apply as has been expected. The first result is
more promising from the point of view of implementing quantum computers of a
useful size, for example with $N\thickapprox 10^{4}$. In the general case
where the qubits vibrate collectively, the evaluation is only approximate
however and a full treatment will involve determining the matrix elements $%
S_{i\alpha ;K}$ relating qubit CM displacements to the vibrational normal
coordinates in order to examine issues such as whether the qubits decohere
independently or collectively.

\section{Summary}

Decoherence effects in quantum computers have been studied for the situation
where the number of qubits $N$ becomes large. A standard model involving $N$
two state qubit systems was treated, with localised, well-separated qubits
undergoing vibrational motion in the trapping potentials. Coherent one and
two qubit gating processes were controlled by time dependent localised
classical EM and magnetic fields, the two qubit gating processes being
facilitated by a cavity mode ancilla. The qubits were coupled to bath of
spontaneous emission modes, the cavity mode was coupled to a bath of cavity
decay modes. The numerous vibrational qubit modes also behave as a
reservoir. Non-RWA couplings and effects of qubit vibrational motion were
both included. The predominant coupling of qubits to the baths is amplitude
coupling via the $\sigma _{X}^{i}$ Pauli operators. Decoherence effects were
specified by the fidelity, and non-Markovian expressions for the short time
behaviour of the fidelity were obtained for the general case where the
qubits and ancilla are in any pure state and the baths and qubit vibrational
modes are in thermal states.

Characteristic decoherence time scales were evaluated for specific qubit
states in the simple situation when no gating processes are occurring and at
zero temperature. The decoherence time is mainly affected by the spontaneous
emission decay of the qubits. Decoherence times scaling inversely as $1/%
\sqrt{N}${\small \ }were found for the cases treated. For the uncorrelated
Hadamard state a decoherence time scale of about $10^{-8}$s for $%
N\thickapprox 10^{4}$\ qubits was found, for a macroscopic $N\thickapprox
10^{22}$\ it was about $10^{-17}$s. For the correlated GHZ state a
decoherence time scale of about $10^{-17}$s for $N\thickapprox 10^{4}$\
qubits was found, for a macroscopic size quantum computer with $%
N\thickapprox 10^{22}$\ it was about $10^{-26}$s. Correlated states
generally were found to have similar decoherence times to the GHZ state, and
the qubits were found to decohere independently.

Characteristic decoherence time scales were also evaluated for specific
qubit states in a further simple situation when no gating processes are
occurring, the temperature is zero, and decay via spontaneous emission modes
is ignored. The decoherence time is now mainly affected by the Lamb-Dicke
coupling of the qubits to the cavity mode ancilla and to the CM\ vibrational
modes. Decoherence times scaling inversely as $1/\sqrt{N}${\small \ }for the
cases treated were found. A decoherence time scale of about $10^{-8}$s for $%
N\thickapprox 10^{4}$\ qubits was found, for a macroscopic size quantum
computer with $N\thickapprox 10^{22}$\ it was about $10^{-17}$s.

The case where spontaneous emission decay can be disregarded, such as when
the qubits are all in a very high Q cavity, is far more promising from the
point of view of implementing quantum computers of useful size, say with $%
N\thickapprox 10^{4}$\ qubits. An alternative, though more complex model for
avoiding spontaneous emission effects would be to replace the two internal
state qubits by qubits with three internal states, such as in a lambda
system. In such a model the qubit states would be two near degenerate ground
states, and the upper state would be in effect part of the ancilla system
and only coming into play during the gating processes. Spontaneous emission
would then only be important during the gating processes and could be
minimized if the latter occurred in short enough times. A similar model has
been examined in a different context in \cite{Beige00a}.

Finally, although the decoherence times found here are not long compared to
bath correlation times, and hence non-Markovian conditions apply as
expected, it would be of interest to consider the behaviour of the fidelity
in the Markovian regime for the present model in order that the short time
scale results for decoherence effects presented here can be linked up with
those for longer time scales.

\bigskip

\section{Figure caption}

Figure 1. Model of an $N$ qubit quantum computer. Two state qubits are
localised around well-separated positions via trapping potentials, and
undergo centre of mass (CM) vibrational motions. Coherent one and two qubit
gating processes are controlled by time dependent localised classical
electromagnetic (EM) fields and magnetic fields that address specific
qubits. Two qubit gating processes are facilitated by a cavity mode ancilla,
which permits state interchange between qubits. Magnetic fields are used to
bring specific qubits into resonance with the classical EM fields or the
cavity mode. The two state qubits are coupled to a bath of EM field
spontaneous emission (SE) modes, and the cavity mode is coupled to a bath of
cavity decay modes. For large $N$ the numerous vibrational modes of the
qubits also act as a reservoir, coupled to the qubits, the cavity mode and
the SE modes.

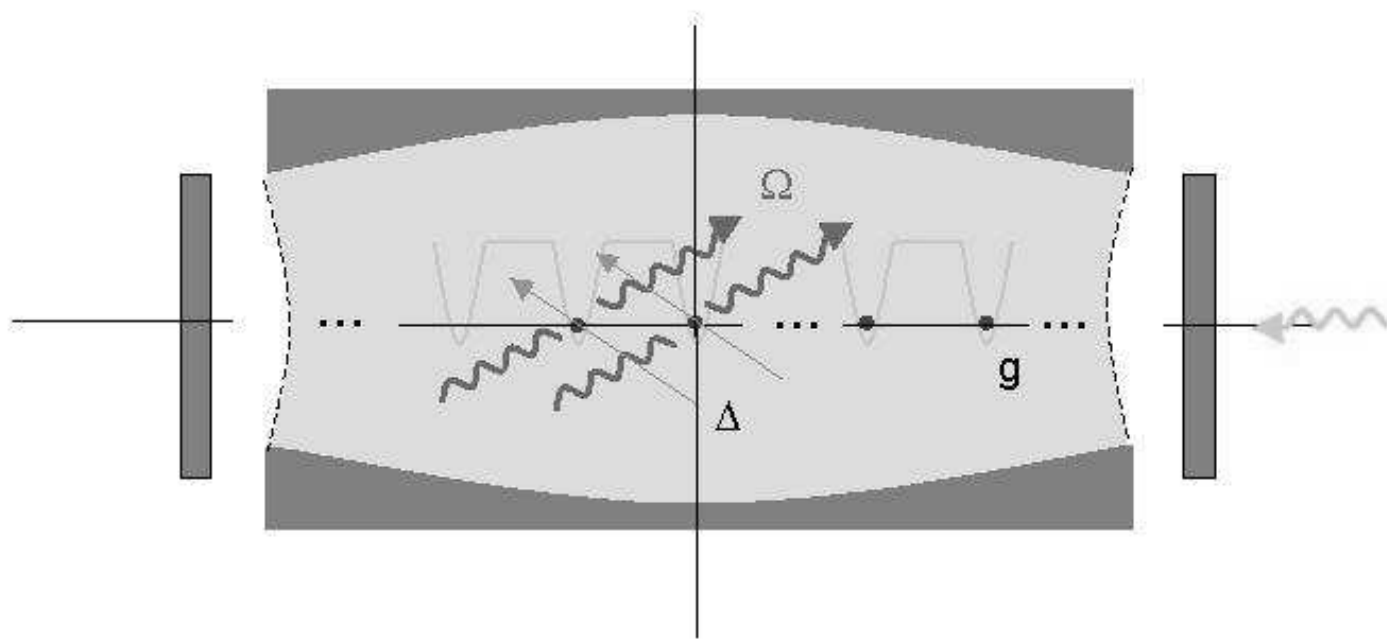
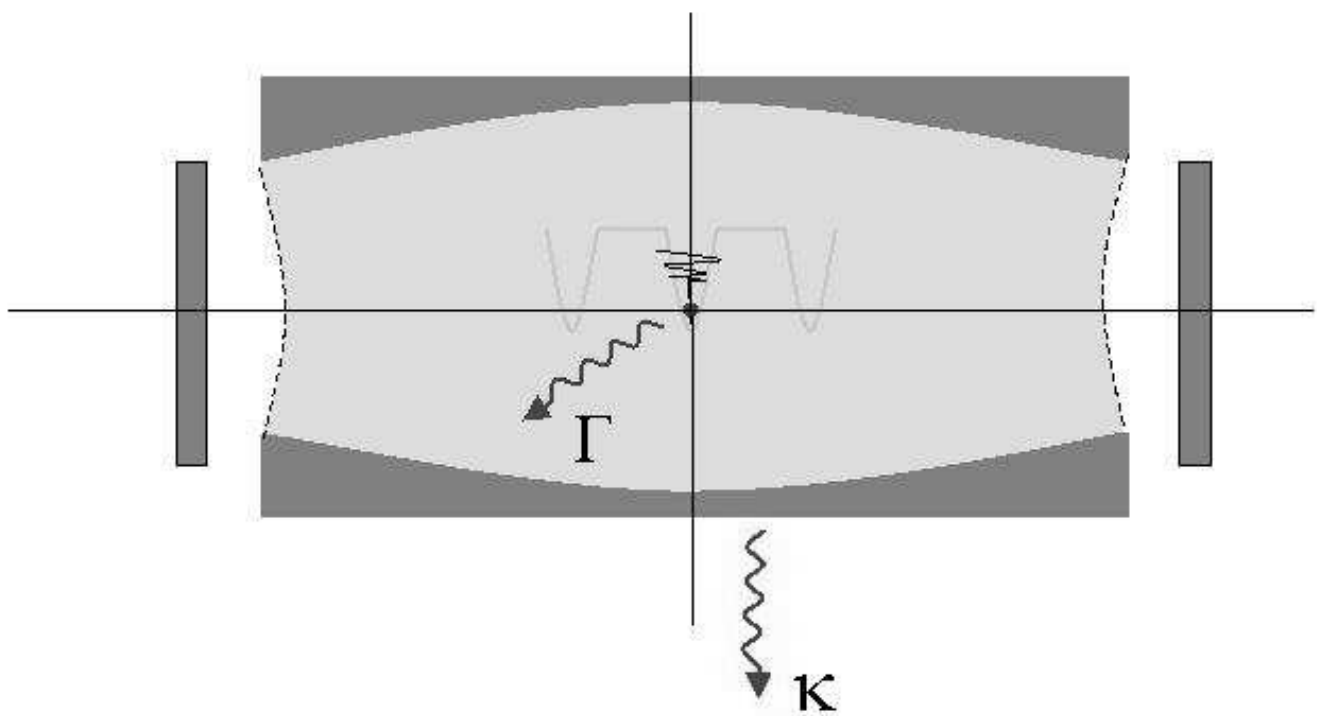
\end{document}